\documentclass[amsmath,amssymb,14pt]{article}

\usepackage{graphicx, epsfig}

\usepackage{amssymb,amsmath,color}

\usepackage{subcaption}
\usepackage{caption}
\usepackage[export]{adjustbox}

\usepackage{dcolumn}
\usepackage{bm}
\newcommand{\be}{\begin{equation}}
\newcommand{\ee}{\end{equation}}
\newcommand{\beqs}{\begin{eqnarray}}
\newcommand{\eeqs}{\end{eqnarray}}

\begin{document}
\title{\textsc{
Green's Function Formulation of Multiple Nonlinear Dirac $\delta$-Function Potential in One Dimension} }

\author{\centerline {Fatih Erman$^1$, Haydar Uncu$^{1,2}$}
\\\and
 {\scriptsize{$^1$
Department of Mathematics, \.{I}zmir Institute of Technology, Urla,
35430, \.{I}zmir, Turkey}}
\\\and
 {\scriptsize{$^2$
Department of Physics, Adnan Menderes University, Aydın, Turkey}} 
\\
{\scriptsize{E-mail: fatih.erman@gmail, huncu@adu.edu.tr}}}

\maketitle

\begin{abstract}

In this work, we study the scattering problem of the general nonlinear finitely many Dirac delta potentials with complex coupling constants (or opacities in the context of optics) using the Green's function method and then find the bound state energies and the wave functions for the particular form of the nonlinearity in the case of positive real coupling constants.   

\end{abstract}


\section{Introduction}

Dirac $\delta$ potential, a particular class of point interactions \cite{Albeverio}, has been considered as an exactly solvable toy model in one-dimensional or effectively one dimensional quantum mechanics in the sense that one can explicitly find the bound state and scattering state solutions. Such potentials are studied for a long time in various areas of physics \cite{Albeverio, Demkov}, where de-Broglie wavelength of a particle is large compared to the spatial extension of an interaction. This subject is still an active area of research, see e.g. \cite{ZTZ}.

The applications of Dirac $\delta$ potential is not only limited to the phenomena in the linear theory of quantum mechanics. A nonlinear version of the Schr\"{o}dinger equation \cite{Moya, Molina}
\begin{equation}
    -{d^2 \psi(x) \over d x^2} - \Omega \delta(x) |\psi(x)|^{\alpha} \psi(x) = E \psi(x) \;,
\end{equation}
describes the optical wave propagation in a one-dimensional linear medium containing a narrow Kerr-type nonlinear strip within the scalar approximation. Here $\Omega$ is called the opacity and $\alpha$ is the nonlinear exponent associated with the Dirac delta potential. In a recent study, the authors \cite{SM16} have found the exact solution for the scattering of the plane wave in the linear medium on the nonlinear $\delta$ function potential and the splitting of the incident solitons is studied perturbatively. This model has been proposed as a soliton-based interferometer \cite{SM16}.
The possible generalization of the above model includes periodic and
quasiperiodic arrays of nonlinear strips for modeling wave propagation in some nonlinear superlattices \cite{HTMG94, HGTM94, SHMT94, HT99}. Moreover, the same form of the above equation is also a useful model in many body quantum systems, where the interacting $N$-body problem
is reduced to an effective noninteracting system using mean-field approximation. Within this approach, the equation for $\alpha=2$ effectively describes an electron propagating in a one-dimensional linear medium that contains a vibrational impurity at the origin that can interact strongly with the electron \cite{CMT93}.  
This simple nonlinear
equation is also important in illustrating the so-called bistability in optics without the need of a feedback mechanism. 
In \cite{RDH}, a similar model including the linear Dirac delta potentials has been studied by treating the nonlinearity perturbatively. 
The discrete version of the above equation is also known as the discrete nonlinear Schr\"{o}dinger equation (DNLS), which was introduced
in the polaron problem in condensed matter
physics \cite{Holstein}. To the best of our knowledge, the nonlinear Dirac delta function in a stationary problem was first introduced in the context of the discrete nonlinear Schrodinger equation in \cite{MT93}.

In this paper, we generalize the scattering problem for finitely many nonlinear Dirac $\delta$ potentials in one dimension with arbitrary complex opacities (coupling constants in the quantum case) using the Green's function method. The main advantage of the method is that the function $\psi$ at the location of the delta centers can be found recursively so that the transmission and the reflection amplitudes are obtained more systematically compared to the methods formerly introduced in the literature \cite{Moya, Molina, Ali1}. We show that this method is useful to find the bound states energies and the wave functions for one \cite{Molina} and double nonlinear  Dirac $\delta$ potential  with  real coupling constants in one dimension as well. Throughout this work, we will assume that our system consists of a particle or an optical wave
propagating through a medium that contains  nonlinear Dirac $\delta$-function potentials.

The paper is organized as follows: In Section \ref{Scattering Problem for a Single Nonlinear Dirac Delta Potential} we solve the scattering problem for a single general nonlinear Dirac $\delta$ potential with complex opacity using the Green' s function method. Section \ref{Scattering Problem for Nonlinear Double delta Function Potential} contains the discussion of the same scattering problem for double nonlinear Dirac $\delta$ potential. In section \ref{Scattering Problem for Nonlinear Multiple Dirac Delta Function Potential}, we generalize the results obtained in the previous sections to finitely many Dirac $\delta$ centers. In section \ref{Bound State Problem for a Single Nonlinear Dirac Delta Function Potential}, we investigate the bound state problem of a single and double nonlinear Dirac $\delta$ potential with real coupling constants.  Finally, the main results are summarized in the conclusion.

\section{The Scattering Problem for a Single Nonlinear Dirac Delta Potential} \label{Scattering Problem for a Single Nonlinear Dirac Delta Potential}

We consider the ``stationary" non-linear Schr\"{o}dinger equation with the most general nonlinear single delta-function potential given in the following form:

\begin{equation} \label{NLS general 1delta}
-{d^2 \psi(x) \over d x^2} + \delta(x-c) f(|\psi(x)|) \psi(x) = k^2 \psi(x) \;,
\end{equation} 
where $\psi(x)$ is in general a complex-valued function and $f(|\psi(x)|)$ is a continuous function, and $k$ denotes the wave number. The case for the constant $f$ corresponds to the well-known linear Dirac delta function potential problem in quantum mechanics.

This problem is actually equivalent to the linear problem with the effective opacity $f_i(|\psi(c_i)|)$ due to the property of Dirac delta functions. However, as we will see, the solutions depend on the unknown $\psi(c_i)$'s that must be found. This gives rise to all the interesting nonlinear effects, e.g., bistability, modulational instability, etc \cite{Moya, Malomed}.

If we rewrite the above equation by leaving the interaction term alone,
we formally obtain an integral equation for the scattering solution associated with the left incident wave:
\begin{equation} \label{LS psil 1delta}
\psi_{l}(x)= A^l e^{i k x} + \int_{-\infty}^{\infty} G^{(+)}(x-x') \delta(x'-c) f^l(|\psi_l(x')|) \psi_l(x') \; dx' \;,
\end{equation}
where $G^{(+)}$ is the Green's function for the Helmholtz equation with outgoing boundary condition in one dimension, given by \cite{Stakgold2}
\begin{equation}
G^{(+)}(x-x')=-{i \over 2 k} e^{i k |x-x'|} \;.
\end{equation}
The equation (\ref{LS psil 1delta}) is also known as the Lippmann-Schwinger equation. Thanks to the Dirac delta function, we find the scattering solution for the left incident wave in terms of the unknown $\psi_l(c)$
\begin{equation}
\psi_l(x)=A^l e^{i k x} -{i \over 2 k} e^{i k |x-c|} f^l(|\psi_l(c)|) \psi_l(c) \;. \label{presolpsil}
\end{equation}
By evaluating this for $x=c$, we obtain the following consistency condition
\begin{equation}
\psi_l(c)={A^l \over \left(1+ {i \over 2k} f^l(|\psi_l(c)|) \right)} e^{i k c} \;. \label{psil(c)}
\end{equation} 
Substituting this into the solution (\ref{presolpsil}) for the left incident wave, we obtain
\begin{eqnarray}
 \psi_l(x) =
  \begin{cases}
 A^{l} \left( e^{i k x} - {i f^l e^{2ik c} \over 2k+i f^l} \; e^{-ik x} \right)  & \text{for} \; x<c \\ \\
 A^{l} e^{i k x} \left( 1 - {i f^l \over 2k+i f^l} \right)   & \text{for} \; x>c \;,
  \end{cases} 
\end{eqnarray}
from which we can easily read off the left reflection and the left transmission amplitudes, 
\begin{eqnarray}
R^l &=& {- i e^{2ikc} f^l \over 2k+ i f^l} = -{g^l \over 1+g^l} \;  e^{2ikc} \;, \label{Reflection amplitudes for 1 delta} \\
T^l & = & {2k \over 2k +if^l} = {1 \over 1+g^l}\;, \label{Transmission amplitudes for 1 delta}
\end{eqnarray}
respectively. Here we have defined $g^l = {i \over 2k} f^l$ for simplicity. These results are consistent with the one obtained by the transfer matrix and nonlinear scattering formulation using the Jost solutions in \cite{Ali1}. However, the expressions for the left reflection and the left transmission amplitudes include the unknown factor $\psi_{l}(c)$. In order to find it, we need to solve the consistency condition for $\psi_l(c)$. For this purpose, 
we take the absolute value of both sides of the equation (\ref{psil(c)}) and show that $|\psi_l(c)|$ is a solution of the following equation
\begin{equation}
x^2 |\hat{f}^{l}|^2 - 2 x^2 \mathtt{Im}(\hat{f}^{l}) + x^2-|A^l|^2 =0 \;, \label{equation for rl}
\end{equation}
where $\hat{f}^{l}:={f^l \over 2k}$. Actually this equation is exactly the same expression with the equation (29) in \cite{Ali1}, where $|\psi_l(c)|$ is replaced by $|N_l|$. This correspondence is consistent with the definition of $N_l$, given by the initial condition $\psi_l(c)=N_l e^{ik c}$ in \cite{Ali1}. 
For the choice of localized Kerr nonlinearity, 
\begin{equation}
    f^l(|\psi_l(c)|)=z (|\psi_l(c)|)^2 \;,
\end{equation}
the above equation (\ref{equation for rl}) is simplified to
\begin{equation}
|\hat{z}|^2 x^6 - 2 \mathtt{Im}(\hat{z}) x^4 + x^2-|A^l|^2 =0 \;, \label{N_l equation for 1 delta}
\end{equation}
where $\hat{z}=z/2k$ and it is a cubic equation in $x^2$ and only one of them is positive and real, and we call it $|\psi_l(c)|^2$. This result tells us that $|\psi_l(c)|$ is a single-valued function of $A^l$, so that the reflection and the transmission amplitudes are uniquely determined, as emphasized in \cite{Ali1}.

The final step for the scattering problem from a single nonlinear Dirac $\delta$ potential is  to find $|\psi_l(c)|$ in terms of $|A^{l}|$ by solving the equation (\ref{N_l equation for 1 delta}) and then substitute these into the reflection and the transmission amplitudes given by (\ref{Reflection amplitudes for 1 delta}) and \eqref{Transmission amplitudes for 1 delta}, respectively.

For the right incident wave, we can similarly obtain the same results for right reflection amplitude $R^r$ and the right transmission amplitude $T^r$ with the one given in  \cite{Ali1} by starting with 
\begin{equation}
\psi_r(x)=A_r e^{-ikx} + \int_{-\infty}^{\infty} G^{(+)}(x-x') \delta(x'-c) f^r(|\psi_r(x')|) \psi_r(x') \; dx' \;,
\end{equation} 
and obtain the same form of the equation (\ref{equation for rl}) for $|\psi_r(c)|$.

\section{The Scattering Problem for Nonlinear Double Dirac Delta Function Potential} 
\label{Scattering Problem for Nonlinear Double delta Function Potential}

We consider now the scattering problem of two nonlinear delta function potential 
\begin{equation} \label{NLS general 2delta}
-{d^2 \psi(x) \over d x^2} + \delta(x-c_1) f_1(|\psi(x)|) \psi(x) + \delta(x-c_2) f_2(|\psi(x)|) \psi(x) = k^2 \psi(x) \;,
\end{equation} 
where $c_2>c_1$ without loss of generality. We apply the same argument again, with the single delta term replaced by the sum of the delta terms, to obtain the solution for the left incident wave
\begin{equation} \label{psil2delta}
\psi_l(x)=A^l e^{ik x} -e^{ik|x-c_1|}g_{1}^{l} \psi_l(c_1)-e^{ik|x-c_2|}g_{2}^{l} \psi_l(c_2) \;, 
\end{equation}
where 
\begin{equation}
g^{l}_i:= g^{l}_i(\vert \psi_l(c_i) \vert) = {i \over 2k} f^{l}_{i}(|\psi_l(c_i)|)    \; .
\end{equation}
Evaluating the formal solution (\ref{psil2delta}) at $x=c_1$ and $x=c_2$, we get the following consistency conditions
\begin{eqnarray}
(1+g^{l}_1) \psi_l(c_1) + g^{l}_2 e^{ik(c_2-c_1)} \psi_l(c_2) & = & A^l e^{i k c_1} \;, \\
g^{l}_1 e^{ik(c_2-c_1)} \psi_l(c_1) + (1+g^{l}_2) \psi_l(c_2) & = & A^l e^{i k c_2} \;.
\end{eqnarray}
This can be viewed as a linear system of equations in terms of $\psi_l(c_1)$ and $\psi_l(c_2)$, so the formal solutions can be easily obtained:
\begin{eqnarray}
\psi_l(c_1) & = & {A^l \over \det \Phi^l} \left[ (1+g^{l}_2) e^{ik c_1}- g^{l}_2 e^{ik(2c_2-c_1)}\right] \;, \label{psilc1} \\
\psi_l(c_2) & = & {A^l \over \det \Phi^l} e^{i k c_2} \;, \label{psilc2}
\end{eqnarray}
where we have defined the matrix $\Phi$ as
\begin{equation}
\Phi^l= 
 \begin{pmatrix}
(1+g^{l}_1) & g^{l}_2 e^{ik(c_2-c_1)} \\ \\
g^{l}_1 e^{ik(c_2-c_1)} & (1+g^{l}_2)  \\
\end{pmatrix} \;.
\end{equation}
Then, substituting the solutions $\psi_l(c_1)$ and $\psi_l(c_2)$ into the equation (\ref{psil2delta}), we finally obtain the scattering solution  
\begin{eqnarray}
 \psi_l(x) =
  \begin{cases}
 A^{l} \left( e^{i k x} - {e^{2ik c_1} \left[g^{l}_{1} + e^{2ik(c_2-c_1)} g_2 +(1-e^{2ik(c_2-c_1)} )g^{l}_{1} g^{l}_{2}  \right] \over \det \Phi^l} \; e^{-ik x} \right)  & \text{for} \; x<c_1 \;, \\ \\ 
 {A^{l} \over \det \Phi^l} e^{i k x}    & \text{for} \; x>c_2 \;.
  \end{cases} 
\end{eqnarray}
Hence, we can easily find the left reflection and the left transmission amplitudes from this solution
\begin{equation}
R^l  = - {e^{2ik c_1} \left[g^{l}_{1} + e^{2ik(c_2-c_1)} g_2 +(1-e^{2ik(c_2-c_1)})  g^{l}_{1} g^{l}_{2} \right] \over \det \Phi^l}  \;,
\end{equation}
and
\begin{equation}
T^l  =  {1 \over \det \Phi^l} \;, \label{TrAmpN2}   
\end{equation}
which are in agreement with the ones given in \cite{Ali1}.

As one may notice that the left reflection and the left transmission amplitudes obtained above are expressed in terms of the unknown quantities $g_{1}^{l}$ and $g_{2}^{l}$, which are functions of $|\psi_{l}(c_i)|$'s for the given form of $f^l$. We are then left with the task of determining these unknowns. For this purpose, we take the absolute value of the solution $\psi_l(c_2)$ in the equation (\ref{psilc2}), and find that  
\begin{equation}
|\psi_l(c_2)| |(1-e^{2ik(c_2-c_1)}) g^{l}_{1} g^{l}_{2} + g^{l}_{1} + g^{l}_{2} +1| -|A^l|=0 \;. \label{equation for psil2 delta}
\end{equation}
This is exactly the same equation obtained by the transfer matrix technique \cite{Ali1}. It is worth pointing out that the equation (\ref{equation for psil2 delta}) allows us to determine $|\psi_l(c_2)|$ due to the fact that we can express $\psi_l(c_1)$ solely in terms of $\psi_l(c_2)$, by simply taking the absolute value of the equation (\ref{psilc1}) and using the result (\ref{psilc2})
\begin{equation}
|\psi_l(c_1)|=|\psi_l(c_2)| |g^{l}_{2}(1-e^{2ik(c_2-c_1)})+1| \;.
\end{equation}
Similarly, one can also easily find the right reflection and the right transmission amplitudes by following the same line of arguments above.

The only point remaining concerns the behaviour of $|R^{l/r}|^2$ and $|T^{l/r}|^2$ by choosing particular form of nonlinearity, such as $f_i(x)=z_i x^{\alpha_i}$. The numerical results for different choices of nonlinear exponents and opacities are explicitly discussed in a recent paper \cite{Ali1} and studied for the case of real values of opacities in \cite{Moya}.
Since our aim is to solve the problem using the Green's function formalism, we are not going to present these numerical results and discussions for double Dirac delta centers given in the above cited works.

\section{The Scattering Problem for Nonlinear Multiple Dirac Delta Function Potential}

\label{Scattering Problem for Nonlinear Multiple Dirac Delta Function Potential}

The main advantage of our formulation for the scattering problem compared with the other methods (such as the transfer-matrix approach and the recently introduced nonlinear scattering approach using Jost functions \cite{Ali1}) for solving nonlinear $\delta$ potential scattering  is that it is possible to solve directly the nonlinear multiple $\delta$ function potential in a systematic way. In order to illustrate this, let us consider 
\begin{equation} \label{NLS general Ndelta}
-{d^2 \psi(x) \over d x^2} + \sum_{i=1}^{N} \delta(x-c_i) f_i(|\psi(x)|) \psi(x)  = k^2 \psi(x) \;.
\end{equation} 
The formal solution for a left incident wave is given by
\begin{equation}
\psi_l(x)=A^l e^{ikx} -{i \over 2k} \sum_{i=1}^{N} e^{ik|x-c_i|} f_i(|\psi_l(c_i)|) \psi_l(c_i) \; . \label{psilNdelta}
\end{equation}
Evaluating this at $x=c_i$, we get the following consistency condition written in a matrix form:
\begin{equation}
\sum_{j=1}^{N} \Phi^l_{ij} \psi_l(c_j)= A^l e^{ikc_i} \;, \label{matrixeq}
\end{equation}
where the matrix $\Phi^l$ is defined by
\begin{eqnarray}
\Phi^l_{ij} = \begin{cases}
1+ {i \over 2k} f_i(|\psi_l(c_i) |) & \text{if} \; i=j \\ \\
   {i \over 2k} f_j(|\psi_l(c_j)|) e^{ik|c_i-c_j|} & \text{if} \; i \neq j \;.  \end{cases}  
\label{phigeneral}   
\end{eqnarray}
So we have transformed the mathematics of the problem from the solution of a  differential equation to an algebraic problem which turns out to be useful  for the general finite number of $\delta$ center case.  In order to explain our solution strategy, we first present the scattering problem from three non linear $\delta$ centers. Then, the equation \eqref{matrixeq} for $N=3$ becomes,
\begin{equation}
 \begin{pmatrix}
(1+g^{l}_1) & g^{l}_2 e^{ik(c_2-c_1)} & g^{l}_3 e^{ik(c_3-c_1)} \\ \\
g^{l}_1 e^{ik(c_2-c_1)} & (1+g^{l}_2)   & g^{l}_3 e^{ik(c_3-c_2)} \\ \\
g^{l}_1 e^{ik(c_3-c_1)} &  g^{l}_2 e^{ik(c_3-c_2)}   & (1+ g^{l}_3 ) 
\end{pmatrix} \; \begin{pmatrix}
\psi_l(c_1) \\ \\
\psi_l(c_2) \\ \\
\psi_l(c_3) 
\end{pmatrix}  =  \begin{pmatrix}
A^l e^{ikc_1} \\ \\
A^l e^{ikc_2} \\ \\
A^l e^{ikc_3} 
\end{pmatrix} \; ,
\label{matrixeq3}
\end{equation}
where we take $c_1<c_2<c_3$ without loss of generality. We first multiply each $i^{th}$ row of the above 
system of equations by $e^{-i k c_i}$ and then we write the new linear system of equations in the following augmented matrix form  
\begin{equation}
 \begin{pmatrix}
(1+g^{l}_1)e^{-ikc_1}  & g^{l}_2 e^{ik(c_2-2 c_1)} & g^{l}_3 e^{ik(c_3-2c_1)} & A^l \\ \\
g^{l}_1 e^{-ikc_1} & (1+g^{l}_2) e^{-ikc_2}  & g^{l}_3 e^{ik(c_3-2c_2)} & A^l \\ \\
g^{l}_1 e^{-ikc_1} &  g^{l}_2 e^{-ik c_2}   & (1+ g^{l}_3 ) e^{-ikc_3} &A^l
\end{pmatrix}\; . 
\label{AugMat3}
\end{equation}
In order to reduce the above system to the row echelon form, we first 
change the first row (R1) and the second row (R2) by subtracting the third row (R3) from them; then change the third row by subtracting $g_{1}^{l}$ times (R1) from it; then replace the third row by subtracting $(1+ g_{1}^{l}+ g_{3}^{l} - g_{1}^{l} g_{3}^{l} (e^{2ik(c_3-c_1)}-1))$ times (R2) from it; and then change the last row by subtracting $e^{-ik c_2} g^{l}_2 \left[ 1- g^{l}_1 (e^{2ik(c_2-c_1)}-1) \right]$ times (R2) from it. As a result of these row reduction operations, we finally get
\begin{equation}
 \begin{pmatrix}
 e^{-ikc_1} & g^{l}_2 e^{-ikc_2} (e^{2ik(c_2-c_1)}-1) &  e^{-ikc_3} \left[-1+ g^{l}_3 (e^{2ik(c_3-c_1)}-1) \right] & 0\\ \\
0 &  e^{-ikc_2}  &  e^{-ik c_3} \left[ -1 + g^{l}_3 (e^{2ik(c_3-c_2)}-1)  \right] & 0 \\ \\
0 & 0   & e^{-ik c_3} \, \det \Phi & A^l   
\end{pmatrix}  \
\; , 
\label{matrixeqfin3}
\end{equation}
where $\det \Phi^l$ is the determinant of the $3 \times  3$ matrix given by the equation \eqref{matrixeq3} and its explicit form is given by 
\begin{eqnarray}
\det \Phi^l &=& 1+ g^{l}_1 + g^{l}_2+g^{l}_3+\left[ g^{l}_1 g^{l}_2 (1-e^{2 ik (c_2-c_1)}) + g^{l}_1 g^{l}_3 (1-e^{2 ik (c_3-c_1)}) + g^{l}_2 g^{l}_3 (1-e^{2 ik (c_3-c_2)}) \right] \nonumber \\
&+& g^{l}_1 g^{l}_2 g^{l}_3 \left[1- e^{2 ik (c_2-c_1)}+e^{2 ik (c_3-c_1)}-e^{2 ik (c_3-c_2)} \right].\label{DetPhTi}
\end{eqnarray}
The appearance of the expression $e^{-ikc_3} \det \Phi^l$ in the last pivot of the reduced matrix \eqref{matrixeqfin3} is actually expected since the reduced matrix is obtained by elementary row operations after scaling each rows of the original system by an exponential factor $e^{-i k c_i}$. Then, the statement is proved once we express the determinant of the $3 \times 3$ partitioned reduced upper triangular matrix $\Tilde{\Phi}$ in terms of the determinant of the matrix $\Phi^l$
\begin{equation}
\det \Tilde{\Phi} = \det \Phi^l \prod_{i}^{3} e^{-ik c_i}  \,  .\label{DetPhTil}
\end{equation}
This equation tells us that we can easily read off $\Tilde{\Phi}_{33}$ without any computation since the determinant of a triangular matrix is the product of its diagonal elements.

The above row reduction procedure helps us to find directly $|\psi_l(c_1)|$, $|\psi_l(c_2)|$, and $|\psi_l(c_3)|$ in terms of each others in such a way that we can easily solve them in a systematic fashion. According to the above reduced matrix \eqref{matrixeqfin3}, we get
\begin{eqnarray}
\psi_l(c_3) &=& \frac{A^l} {\det \Phi} \; e^{ikc_3} \;,
\label{psi3forN3} \\ \cr
\psi_l(c_2) &=& \frac{A^l}{\det \Phi} \; e^{ikc_2} \; \left[1+ g^{l}_3(\vert\psi_l(c_3) \vert ) \left( 1- e^{2ik(c_3-c_2)} \right) \right] \;,    \label{psi2forN3} \\ \cr
\psi_l(c_1) &=& \frac{A^l}{\det \Phi} \; e^{ikc_1} \; \bigg[1+ g^{l}_2 (\vert\psi_l(c_2) \vert )\left( 1- e^{2ik(c_2-c_1)} \right)+g^{l}_3 (\vert\psi_l(c_3) \vert ) \left( 1- e^{-2ik(c_3-c_2)} \right) \cr
 & & \hspace{2cm} +  g^{l}_2 (\vert\psi_l(c_2) \vert ) \, g^{l}_3 (\vert\psi_l(c_3) \vert ) \left( 1- e^{2ik(c_2-c_1)}+ e^{2ik(c_3-c_1)}-e^{2ik(c_3-c_2)} \right) \bigg] \;.  \label{psi1forN3}
\end{eqnarray}
From the above set of equations, one can see that $|\psi_l(c_2)|$ depends only on $|\psi_l(c_3)|$ thanks to the relation \eqref{psi3forN3},
and similarly $|\psi_l(c_1)|$ depends only on $|\psi_l(c_2)|$ and $|\psi_l(c_3)|$. Hence, we can find $|\psi_l(c_3)|$ by taking the absolute value of  the equation \eqref{psi3forN3} and substituting $|\psi_l(c_1)|$ and $|\psi_l(c_2)|$, all expressed in terms of $|\psi_l(c_3)|$. 
Once we determine $|\psi_l(c_3)|$, then $|\psi_l(c_2)|$ and $|\psi_l(c_1)|$ can all be systematically found by just substituting back the value of $|\psi_l(c_3)|$ in the equations \eqref{psi2forN3} and \eqref{psi1forN3}, respectively.   

Finally, it immediately follows from the explicit form of the wave function \eqref{psilNdelta} for $x<c_1$ and $x>c_3$ that  the reflection and the transmission amplitudes become
\begin{eqnarray}
R^l & = & - {e^{2ik c_1} \over \det \Phi^l} \bigg[ g^{l}_{1} + e^{2ik(c_2-c_1)} g^{l}_2 + e^{2ik(c_3-c_1)} g^{l}_3  +(1-e^{2ik(c_2-c_1)} )g^{l}_{1} g^{l}_{2}+ (1-e^{2ik(c_3-c_1)} ) g^{l}_{1} g^{l}_{3} \cr & & +  (1-e^{2ik(c_3-c_2)}) g^{l}_{2} g^{l}_{3} +  g^{l}_1 g^{l}_2 g^{l}_3 \left(1- e^{2 ik (c_2-c_1)}+e^{2 ik (c_3-c_1)}-e^{2 ik (c_3-c_2)} \right) \bigg]  \;,
\\
T^l  &=&   {1 \over \det \Phi^l} \;, \label{TrAmpN3}    
\end{eqnarray}
respectively. 

The important point to note here that the transmission amplitudes $T^l$ given in the equations \eqref{TrAmpN2} and \eqref{TrAmpN3} are of the same  form, namely $1/\det \Phi^l$. Actually, this form can be easily deduced from the fact that the scattering solution $\psi_l$ at the rightmost region ($x>c_3$) is of the form $A^l T ^l e^{i k x}$ as explained in \cite{Ali1}. As a consequence of the continuity of the scattering solution $\psi_l$ at $x=c_3$, this fact implies that $T^l$ for $N=2$ and $N=3$ must be always of the form $1/\det \Phi^l$, where we have used the equations \eqref{psilc2} and \eqref{psi3forN3}. As will be explained in the next paragraph, this form of the transmission amplitude still holds for an arbitrary finite number of delta centers.

\begin{figure}[h!]
\begin{minipage}{5cm}
\includegraphics[scale=0.75]{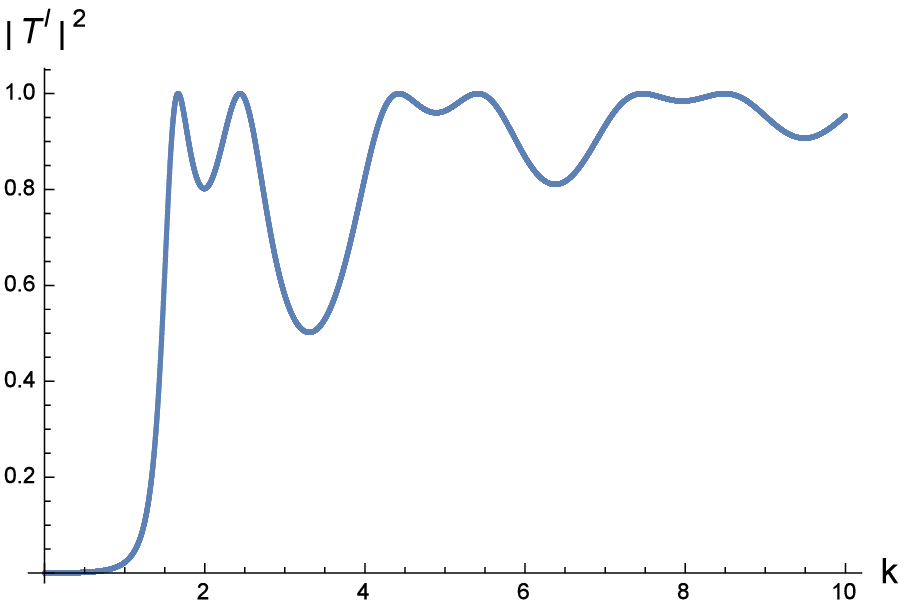}
\end{minipage}
\qquad \qquad \qquad \qquad
\begin{minipage}{5cm}
\includegraphics[scale=0.75]{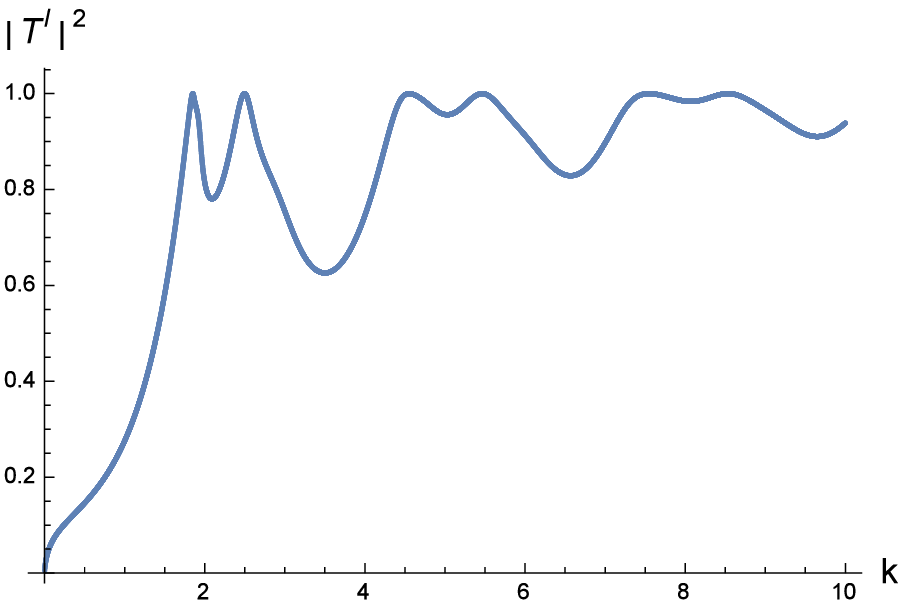}
\end{minipage} \\ \\ \\ \\ \centering
\begin{minipage}{5cm}
\includegraphics[scale=0.75]{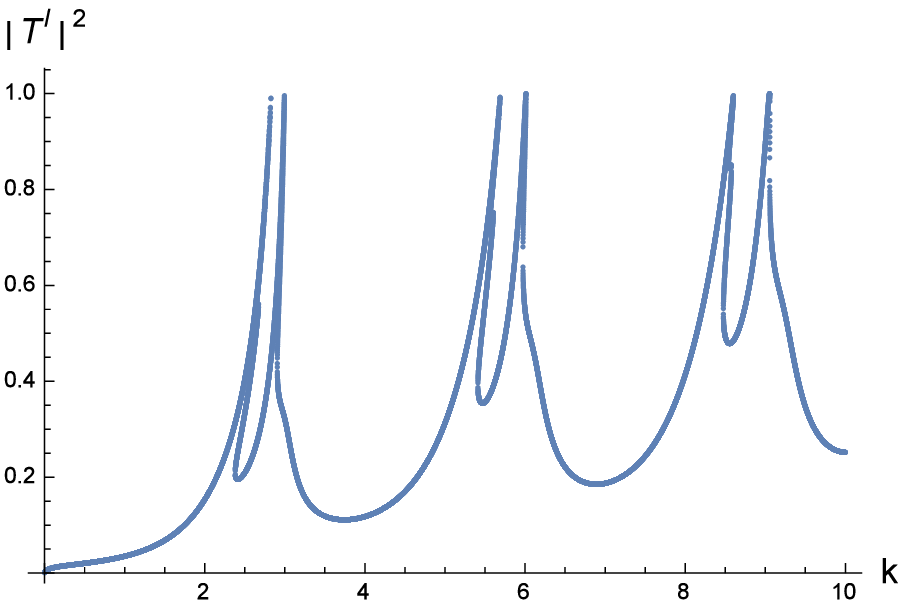}
\end{minipage}
\caption{Transmission intensities $|T^l|^2$ as a function of $k$ for real $z_i$. In all the graphics we take the strengths of all three $\delta$ centers equal to each other, $c_1=0$, $c_2=1$, $c_3=2$ and $|A_l|=1$. The top left figure corresponds to the linear case $\alpha_i=0$ and $z_i= 2$. The top right figure corresponds to weak nonlinear case $\alpha_i=2$ and $z_i= 2$. The below figure corresponds to strong nonlinear case $\alpha_i=2$ and $z_i=20$.} 
\label{realtransmission}
\end{figure}

\begin{figure}[h!]
\begin{minipage}{5cm}
\includegraphics[scale=0.75]{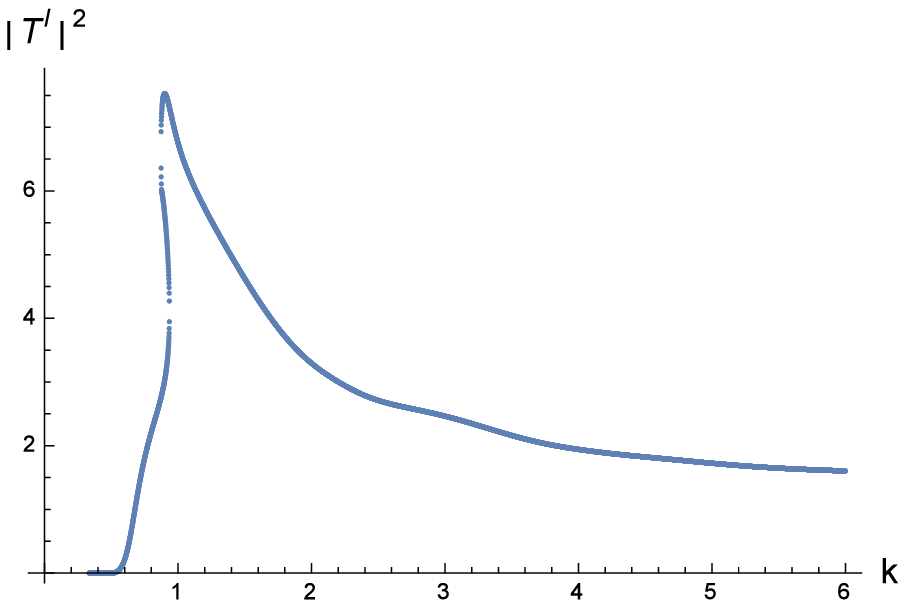}
\end{minipage}
\qquad \qquad \qquad \qquad
\begin{minipage}{5cm}
\includegraphics[scale=0.75]{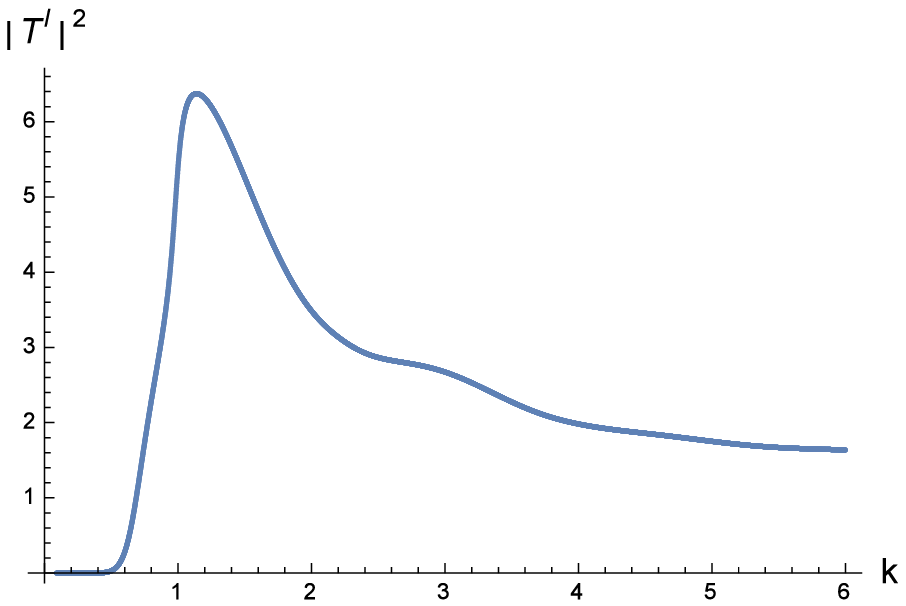}
\end{minipage} \\ \\ \\ \\
\begin{minipage}{5cm} 
\includegraphics[scale=0.75]{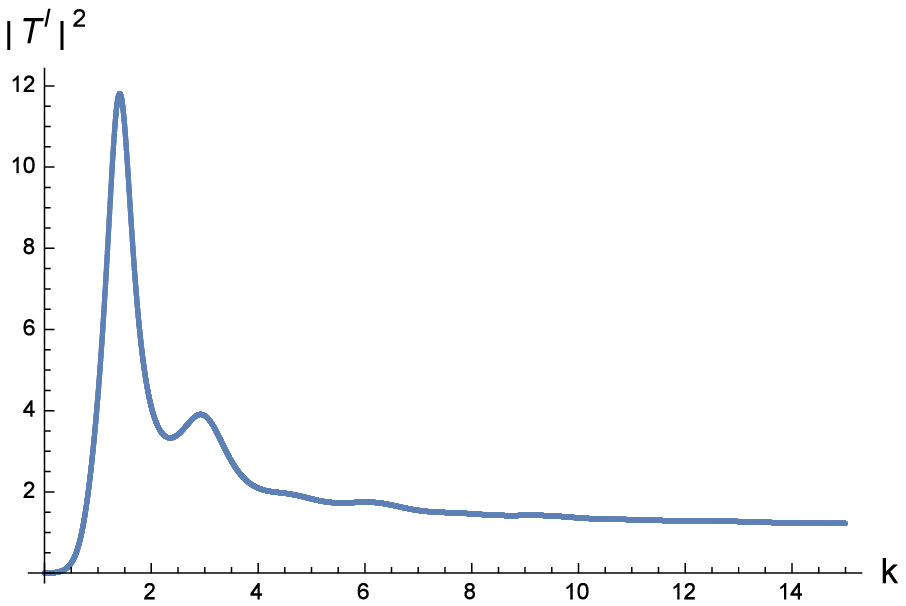}
\end{minipage}
\qquad \qquad \qquad \qquad 
\begin{minipage}{5cm}
\includegraphics[scale=0.75]{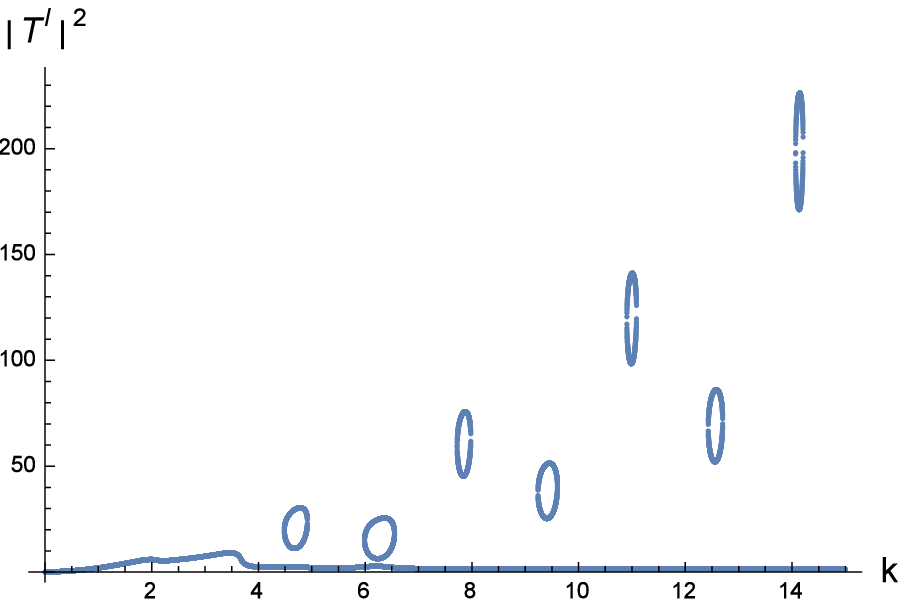}
\end{minipage} \\ \\ \\ \\ \centering
\begin{minipage}{5cm}
\includegraphics[scale=0.75]{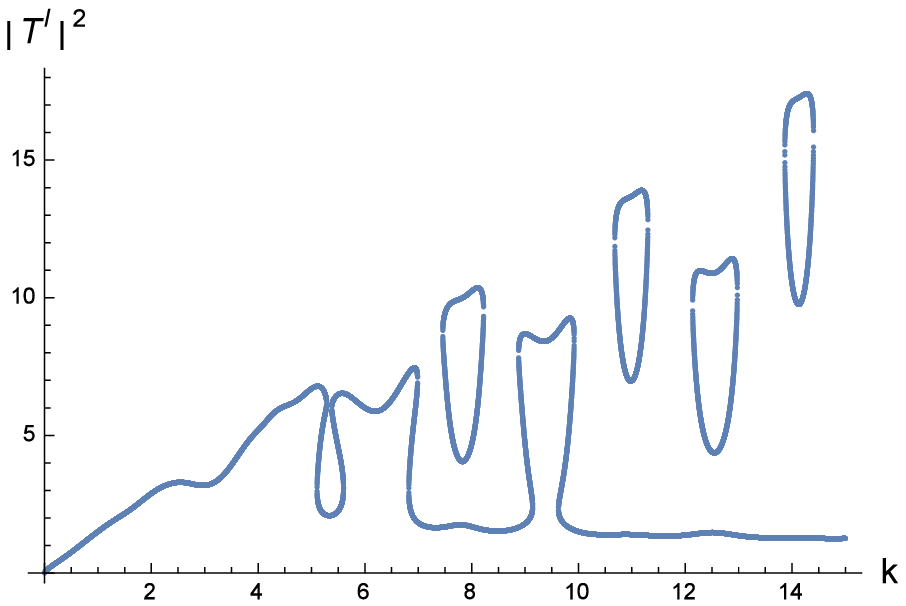}
\end{minipage}
\caption{Transmission intensities $|T^l|^2$ as a function of $k$  for $\mathcal{P}$ symmetric case: $c_1=-1$, $c_2=0$, $c_3=1$. In all the graphics we take all the coupling constants $z_i=i$ and $|A_l|=1$. In the top left graph all $\alpha_i=-0.7$. In the top right graph $\alpha_i=-0.5$. The left graph in the middle row corresponds to the linear case $\alpha_i=0$, in the right graph in the middle row $\alpha_i=1$, and $\alpha_i=2$ in the last graph.} 
\label{imtransmission}
\end{figure}

Now, we plot the graphs of the left transmission intensities  $|T^l|^2$ as a function of $k$ for three $\delta$ function potential. In all the Figures, we choose $f_i(|\psi(x)|)=z_i |\psi(x)|^{\alpha_i}$  in the Equation \eqref{NLS general Ndelta} and choose  all the coupling constants $z_i$ to be equal to each other. In all graphs in the Figure \ref{realtransmission} we show how the left transmission intensities change with respect to wave number $k$ for linear, weak and strong non linear cases, respectively. As in the case of two $\delta$ centers \cite{Moya}, weak nonlinearity slightly changes the transmission intensities compared to linear case. However, the strong nonlinearity causes dramatic effects such as bistability of transmission intensities discussed in \cite{Moya} in detail. We note that the bistabilities occur around the transmission resonances. In the Figure \ref{imtransmission}, we plot the transmission intensities for parity ($\mathcal{P}$) symmetric $\delta$ functions and imaginary coupling constants $z=i$. Therefore the transmission intensities exceed one. In the top graphs, the nonlinear exponents are negative $\alpha_i=-0.7$ and $\alpha_i=-0.5$, respectively. We show the transmission intensity as a function of $k$ for the linear case $\alpha_i=0$. In the right graph of the middle row, $\alpha_i=1$. In the last row of the Figure \ref{imtransmission} we take $\alpha_i=2$. When the coupling coefficients are imaginary and the nonlinear exponents are positive, multistabilities occur for the transmission intensities as can be seen from the last two graphs of the Figure \ref{imtransmission} as in the case of two $\delta$ centers \cite{Ali1}. 

The method we have introduced can be directly generalized to the scattering problem for arbitrarily finitely many nonlinear Dirac $\delta$ centers in the same manner. One of the reason why the above formulation is useful is that one can recursively calculate $\psi_l(c_i)$'s, which is very convenient for numerical computations even if there are more than two delta centers. There is no loss of generality in assuming $c_1<c_2 < \ldots < c_n$ in this case as well. The explicit form of the consistency condition \eqref{matrixeq} becomes
\begin{equation}
 \begin{pmatrix}
(1+g^{l}_1) & g^{l}_2 e^{ik(c_2-c_1)} & \ldots & g^{l}_n e^{ik(c_n-c_1)} \\ \\ 
g^{l}_1 e^{ik(c_2-c_1)} & (1+g^{l}_2) & \ldots  & g^{l}_n e^{ik(c_n-c_2)} \\ \\ 
\vdots & \vdots & \ddots   & \vdots\\ \\ 
g^{l}_1 e^{ik(c_n-c_1)} & g^{l}_2 e^{ik(c_n-c_2)}  & \ldots &  (1+g^{l}_n) 
\end{pmatrix} \; \begin{pmatrix}
\psi_l(c_1) \\ \\
\psi_l(c_2) \\ \\
\vdots \\ \\
\psi_l(c_n) 
\end{pmatrix}  =  \begin{pmatrix}
A^l e^{ikc_1} \\ \\
A^l e^{ikc_2} \\ \\ 
\vdots
\\ \\
A^l e^{ikc_n} 
\end{pmatrix} \; . 
\label{matrixeqN}
\end{equation}
We can now proceed analogously, that is we multiply each row by a corresponding phase i.e., we multiply $i^{th}$ row by $e^{- ikc_i}$. Then we apply the row reduction to the augmented matrix corresponding to the matrix equation \eqref{matrixeqN}
\begin{equation}
 \left( \begin{array}{ccccc}
(1+g^{l}_1)e^{-ikc_1}  & g^{l}_2 e^{ik(c_2-2 c_1)}& \ldots & g^{l}_n e^{ik(c_n-2c_1)} & A^l \\ \\ 
g^{l}_1 e^{-ikc_1} & (1+g^{l}_2) e^{-ikc_2} & \ldots  & g^{l}_n e^{ik(c_n-2c_2)} & A^l \\ \\
\vdots & \vdots & \ddots & \vdots & \vdots \\ \\ 
g^{l}_1 e^{-ikc_1} &  g^{l}_2 e^{-ik c_2} & \ldots   & (1+ g^{l}_n )  e^{-ikc_n} & A^l
\end{array} \right) \; . 
\label{AugMatN}
\end{equation}
We first change all the rows except the last one by subtracting the last row from all the other ones; then change the last row by subtracting $g_{1}^{l}$ times (R1) from it, the first column of the new augmented matrix becomes 
\begin{equation}
    \begin{pmatrix}
    e^{-i k c_1} \\ 0 \\ \vdots \\ 0 
\end{pmatrix}    \;. 
\end{equation}
We continue in this fashion up to $N^{th}$ column in such a way that all the pivots are of the form $e^{-i k c_1}, e^{-i k c_1}, \ldots, e^{-i k c_{n-1}}$ so we get  
\begin{equation}
 \left( \begin{array}{ccccc}
e^{-ikc_1}  & g^{l}_2 e^{ik(c_2-2 c_1)}& \ldots & e^{-ik(c_n)}\left[ -1 + g^{l}_n (e^{2ik(c_n-c_1)}-1)  \right] & 0 \\ \\
0 & e^{-ikc_2} & \ldots  &e^{-ik(c_n)}\left[ -1 + g^{l}_n (e^{2ik(c_n-c_1)}-1)  \right] & 0 \\ \\
\vdots & \vdots & \ddots & \vdots & \vdots \\ \\
0 &  0  & \ldots   & e^{-ikc_n} \textrm{det} \Phi & A^l
\end{array} \right) \; . 
\label{AugMatNF}
\end{equation}
Thus we can immediately write $\psi_l(c_N)$:
\begin{equation}
\psi_l(c_N) = \frac{A^l e^{ikc_N}}{\det \Phi}.
\label{psi3forNN}
\end{equation}
Since the matrix in the equation \eqref{AugMatNF} is an upper triangular matrix, $\psi_l(c_i)$  depends on $\psi_l(c_j)$ only if $j>i$. So being determined $\psi_l(c_N)$ we can determine $\psi_l(c_{N-1})$, then knowing $\psi_l(c_N)$ and $\psi_l(c_{N-1})$ we can determine $\psi_l(c_{N-2})$, and continuing the process we find all the $\psi_l(c_i)$s. Thus all the unknowns in Equation \eqref{psilNdelta} are determined and therefore it is possible to  read off the transmission and the reflection amplitudes from the expression of the wave function given in the Equation \eqref{psilNdelta} as we have done for $N=2$ and $N=3$ cases.

In this section, we have only considered three symmetrically located nonlinear Dirac delta potentials and computed the reflection and the transmission amplitudes for left incoming waves, namely $R^l$ and $T^l$. Because of the symmetry, we expect that the amplitudes $R^l$ and $T^l$ would not change for the right incoming waves, that is, $R^l=R^r$ and $T^l= T^r$. However, they do not have to equal for arbitrary locations of delta centers.

\section{The Bound State Problem for Single and Double Nonlinear Dirac Delta Potential}

\label{Bound State Problem for a Single Nonlinear Dirac Delta Function Potential}

We will now consider the quantum mechanical bound state problem, where the nonlinearity is of the form $|\psi(x)|^{\alpha}$:
\begin{equation}
    -\psi''(x)-\Omega \delta(x-c) |\psi(x)|^{\alpha} \psi(x)= E \psi(x) \;. \label{Bound state 1 delta equation}
\end{equation}
Here $\Omega$ is positive and real coupling constant. We assume that the bound state energy is negative so we write $E=-\nu^2$, where $\nu>0$ for simplicity. In this case, the general solution to the above equation (\ref{Bound state 1 delta equation}) is
\begin{equation}
    \psi(x)=-\Omega \int_{-\infty}^{\infty} G(x-x') \delta(x'-c) |\psi(x')|^{\alpha} \psi(x') \; dx' \;,
\end{equation}
where the homogenous solution to the equation (\ref{Bound state 1 delta equation}) is zero, and $G(x-x')$ is the Green's function, given by \cite{Stakgold2}
\begin{equation}
    G(x-x')=-{1 \over 2\nu} \exp(-\nu|x-x'|) \;.
\end{equation}
Hence, we obtain the general solution for the bound state problem
\begin{equation}
    \psi(x)={\Omega \over 2\nu} \exp(-\nu|x-c|) |\psi(c)|^{\alpha} \psi(c) \;. \label{psi delta N=1}
\end{equation}
Then, by evaluating $\psi(x)$ at $x=c$ we get the consistency condition, which implies 
\begin{equation}
    \nu={\Omega \over 2} |\psi(c)|^{\alpha} \;. \label{nu for 1 delta}
\end{equation}
This result gives us the bound state energy in terms of the unknown quantity $\psi(c)$, i.e.,
\begin{equation}
    E=-{\Omega^2 \over 4} |\psi(c)|^{2\alpha} \;.
\end{equation}
From the normalization condition $\int_{-\infty}^{\infty}|\psi(x)|^2 d x =1$, we have the following constraint
\begin{equation}
{\Omega^2 \over 4 \nu^3} |\psi(c)|^{2 \alpha +2} =1 \;. \label{psi(c) 1 delta}
\end{equation}
Substituting the equation (\ref{nu for 1 delta}) into the above, we finally get the unknown $|\psi(c)|$ in terms of $\Omega$ and $\alpha$
\begin{equation}
    |\psi(c)|= \left( {\Omega \over 2}\right)^{1 \over 2-\alpha} \;. 
\end{equation}
Once we have found $|\psi(c)|$, we have the explicit form of the bound state energy 
\begin{equation}
    E=-{\Omega^2 \over 4} \left( {\Omega \over 2}\right)^{2 \alpha \over 2-\alpha} \;,
\end{equation}
and the bound state wave function  
\begin{equation}
    \psi(x)= \left( {\Omega \over 2}\right)^{1 \over 2-\alpha} \exp\left(- \left({\Omega \over 2}\right)^{2 \over 2-\alpha} |x-c|\right) \;. \label{bound state wave function 1 delta}
\end{equation}
This result is consistent with the linear problem when $\alpha=0$ and there is a singularity for $\alpha \rightarrow 2$, as emphasized in \cite{Molina}. It is worth mentioning that the bound state energies here are found by imposing the normalization condition for the wavefunction, in contrast to the linear case. Actually, one can solve the bound state problem with a general nonlinear term $f(|\psi(x)|)$ by following the same line of arguments except for the fact that we need the explicit form of the nonlinear terms to solve the consistency conditions.

Since the structure of the bound state problem is the same for arbitrary number $N$ of nonlinear Dirac delta centers, we will directly consider the following problem 
\begin{equation}
    -\psi''(x) - \sum_{i=1}^{N} \Omega_i \delta(x-c_i) |\psi(x)|^{\alpha_i} \psi(x) = -\nu^2 \psi(x) \;.
\end{equation}
Then, it immediately follows that the bound state wave function contains the same form of terms as in the equation (\ref{psi delta N=1})  
\begin{equation}
    \psi(x)= \sum_{j=1}^{N} {\Omega_j \over 2\nu} e^{-\nu|x-c_j|}|\psi(c_j)|^{\alpha_j} \psi(c_j) \;.  \label{bound state wave function N delta}
\end{equation}
By evaluating the wave function (\ref{bound state wave function N delta}) at $x=c_i$ and splitting the $j=i$ th term in the sum, we obtain the consistency condition
\begin{eqnarray}
\left(1-{\Omega_i \over 2\nu}|\psi(c_i)|^{\alpha_i}\right)\psi(c_i) - \sum_{\substack{j=1 \\
(j \neq i)}}^{N}  {\Omega_j \over 2\nu} e^{-\nu|c_i-c_j|} |\psi(c_j)|^{\alpha_j} \psi(c_j)=0 \;,
\end{eqnarray}
for all $i$. This can be put in a matrix form
\begin{equation}
\sum_{j=1}^{N}\Phi_{ij} \psi(c_j)=0 \;,
\end{equation}
where 
\begin{eqnarray}
\Phi_{ij}=
\begin{cases}
1-{\Omega_i \over 2\nu}|\psi(c_i)|^{\alpha_i} & \text{for} \; i = j \\
-{\Omega_j \over 2\nu} e^{-\nu|c_i-c_j|} |\psi(c_j)|^{\alpha_j} & \text{for} \; i \neq j \;.
\end{cases}
\end{eqnarray}
For the non-trivial solutions $\psi(c_i)$, we must have
\begin{equation}
    \det \Phi_{ij} = 0 \;, \label{bound state equation}
\end{equation}
from which we can solve $\nu$ (equivalently the bound state energies) in terms of the unknowns $\psi(c_i)$'s and all the parameters in the model. We can find $|\psi(c_i)|$'s from the normalization condition of the wave function (\ref{bound state wave function N delta}) 
\begin{eqnarray}
 \sum_{i=1}^{N} {\Omega_{i}^{2} \over 4 \nu^3} |\psi(c_i)|^{2 \alpha_i}|\psi(c_i)|^2 & + & \sum_{\substack{i,j=1 \\
(i < j)}}^{N} {\Omega_i \Omega_j \over 4 \nu^2} |\psi(c_i)|^{\alpha_i} |\psi(c_j)|^{\alpha_j} \left( \psi(c_i)\psi^{*}(c_j) + \psi^{*}(c_i)\psi(c_j)  \right) \cr & & \hspace{2cm} e^{-\nu(c_j-c_i)} \left({1 \over \nu}  + (c_j-c_i) \right) = 1 \;, \label{consistency N bound}
\end{eqnarray}
where $*$ denotes the complex conjugation. This equation (\ref{consistency N bound}) together with (\ref{bound state equation}) allows us to determine the bound state energies and the bound state wave function in terms of $\alpha_i$ and $\Omega_i$'s. 

In order to get an analytical result, let us consider the double delta centers with equal strengths and nonlinear exponents. Then, the condition (\ref{bound state equation}) reads
\begin{equation}
    \left(2\nu-\Omega |\psi(c_1)|^{\alpha}\right) \left(2\nu-\Omega |\psi(c_2)|^{\alpha}\right) = \Omega^2 e^{-2\nu|c_1-c_2|} \psi(|c_1|)^{\alpha}\psi(|c_2|)^{\alpha} \;. 
\end{equation}
Let $x:=2\nu$, $d:=|c_1-c_2|$, and $\beta_i= \Omega |\psi(c_i)|^{\alpha}$, then the above equation becomes 
\begin{equation}
    (x-\beta_1)(x-\beta_2)=e^{-d x} \beta_1 \beta_2 \;. \label{transcendental}
\end{equation}
This transcendental equation has the same form as in the linear case, so the sufficient condition for two bound states can be easily found as
\begin{equation}
d > {\beta_1 + \beta_2 \over \beta_1 \beta_2} = {|\psi(c_1)|^{\alpha} + |\psi(c_2)|^{\alpha} \over
\Omega |\psi(c_1)|^{\alpha} |\psi(c_2)|^{\alpha}  
} \;.
\end{equation}
In other words, we have at most two bound states (on the left hand side of the equation (\ref{transcendental}) we have a parabola whose zeroes are positive real numbers, whereas the right hand side is a decreasing function of $x$). In contrast to the linear problem, we can not directly solve the above transcendental equation (\ref{transcendental}) analytically, where the solution for the linear case is given in terms of the Lambert W function. Here the problem is more complicated since $\beta_i$'s depend on the value of the wave function at support of the Dirac delta function. Moreover, the solution to the equation (\ref{transcendental}) depends on the unknowns $\psi(c_i)$'s and we have a constraint for these unknowns, given by  the normalization condition  (\ref{consistency N bound}) for $N=2$ with equal strengths and nonlinear exponents    
\begin{eqnarray} & & 
 {\Omega^2 \over 4 \nu^3} \Bigg( |\psi(c_1)|^{2\alpha+2}  +  |\psi(c_2)|^{2\alpha+2} + 2 |\psi(c_1)|^{\alpha}  |\psi(c_2)|^{\alpha} \mathtt{Re}\left( \psi^{*}(c_1) \psi(c_2) \right)  
\cr & & \hspace{3cm}  e^{-\nu(c_2-c_1)} \left(1  + \nu (c_2-c_1) \right) \Bigg) =1 \;. \label{normalization bound state N=2}
\end{eqnarray}
Nevertheless, we can still express the solutions analytically by using the following physical argument. Due to the symmetry of the problem (we can always assume that the double delta potentials are located symmetrically around the origin), we naturally expect that $|\psi(c_1)|=|\psi(c_2)|=A$, so 
\begin{equation}
    \beta_1=\beta_2=\beta=\Omega A^{\alpha} \;. 
\end{equation}
Hence, the equations (\ref{transcendental}) and (\ref{normalization bound state N=2}) are simplified as
\begin{equation}
    (x-\beta) = \pm \beta e^{-{d x \over 2}} \;, \label{trans_simplified}
\end{equation}
and 
\begin{equation}
    A^{2 \alpha + 2} {\Omega^2 \over 2} \left( 1 + \cos\theta \; e^{-{d x \over 2}} \left(1+ {d x \over 2}\right)\right) =1 \;, \label{normalizationsimplified}
\end{equation}
respectively. Here $\theta$ is the relative phase between $\psi(c_1)$ and $\psi(c_2)$, that is, $\psi(c_1)=e^{i \theta} \psi(c_2)$. Moreover, $\cos \theta = 1$ for even parity solution $x=x_+$, whereas $\cos \theta=-1$ for odd parity solution $x=x_{-}$.

Since the solution of the equation $z e^z =y$ is given by the Lambert W function \cite{Lambert}, i.e., $z=W[y]$, the solutions of (\ref{trans_simplified}) are given by 
\begin{equation}
x_{\pm}= \beta_{\pm} + {2 \over d} W \left(\pm {d \beta_{\pm} \over 2} e^{-{d \beta_{\pm} \over 2}}\right) \;, \label{LambertW}
\end{equation}
where $\beta_{\pm}=\Omega A_{\pm}^{\alpha}$. This equation tells how $\nu$ (so the bound state energies) changes with respect to the given parameters $\Omega$, $\alpha$, and the unknown factors $A_{\pm}$. In order to find $A_{\pm}$, we first express the exponential factor $e^{-{d x \over 2}}$ in terms of the linear factor $(x_{\pm}-\beta_{\pm})$ from the equation (\ref{trans_simplified}) and then substitute them into the equation (\ref{normalizationsimplified}), we obtain a single equation both for even and odd parity solutions 
\begin{equation}
4 \beta_{\pm}^2 A_{\pm}^2 \left[ 1 + \left({x_{\pm} \over \beta_{\pm}} -1\right) \left( 1 + {d x_{\pm} \over 2}\right) \right] =x_{\pm}^3 \;. 
\end{equation}
This is a third order polynomial equation in $x_{\pm}$, whose positive explicit solutions can easily be found: 
\begin{eqnarray}
 x_{\pm}= A_{\pm}  \left(\Omega d A_{\pm}^{\alpha+1} \pm \sqrt{\Omega^2 d^2 A_{\pm}^{2 \alpha+2} -2 \Omega A_{\pm}^{\alpha} \left(\Omega d A_{\pm}^{\alpha}-2\right) )}\right)  \;. 
\end{eqnarray}
These solutions must be consistent with the one given by (\ref{LambertW}). Using this fact, we obtain the equations for $A_{\pm}$ so that we can find them in terms of the parameters in the model, namely $\alpha$, $\Omega$, and $d$. Once we have found $A_{\pm}$, we can find the bound state energies and the wavefunctions.

Similar to the linear case, we have realized that there are even and odd parity solutions for the bound states. In reference \cite{MMD08},  the spontaneous symmetry breaking (SSB)  phenomena has been studied and the full analytical solutions corresponding to symmetric, antisymmetric, and asymmetric states trapped by the nonlinear double well potential (including the limiting case, two Dirac delta functions located symmetrically around the origin) are given. We believe that the formalism we have introduced may be helpful in studying the SSB phenomena with symmetrically located nonlinear  multiple or periodic array of Dirac delta function potentials.

\section{Conclusion}
In this paper, we have studied the scattering for finitely many general nonlinear Dirac delta potentials with  complex coupling coefficients using the Green's function. We have first solved the transmission and the reflection amplitudes for single and double Dirac $\delta$ potentials and then have compared our results with the previous ones in the literature. We have seen that the Green's function method allows us to generalize the nonlinear scattering problem, studied previously in \cite{Molina, Ali1} using the transfer matrix method, to finitely many Dirac $\delta$ potentials. Although the transfer matrix method is a useful tool and intuitively easy to understand for scattering problems, we have shown that the Green's function method is a straightforward way of finding the scattering and the bound state problem for nonlinear multiple Dirac $\delta$ centers. Finally, we have illustrated that the Green's functions can also be used to calculate the  bound states for single and double $\delta$ potentials for particular nonlinearity and real coupling constants.

\section{Acknowledgements}
The authors would like to thank A. Mostafazadeh for useful discussions. This work has been done during the visit of HU the \.{I}zmir Institute of Technology as a sabbatical.

\end{document}